\documentclass{ws-ijmpa}
\usepackage[super,compress]{cite}
\usepackage{graphicx}
\usepackage{url}
\def\be{\begin{equation}}
\def\ee{\end{equation}}
\def\ba{\begin{eqnarray}}
\def\ea{\end{eqnarray}}
\newcommand\as{\alpha_S}

\def \gsim{ \mathrel{\vcenter
     {\hbox{$>$}\nointerlineskip\hbox{$\sim$}}}}

\renewcommand\trimmarks{}

\begin{document}
\markboth{Matteo Cacciari}{Phenomenological and theoretical developments in jet physics at the LHC}

%
%

\title{Phenomenological and theoretical developments \\in jet physics at the LHC
\footnote{Contribution to ``Jet Measurements at the LHC'', G. Dissertori ed. To
appear in International Journal of Modern Physics A (IJMPA).}
}

\author{MATTEO CACCIARI
}

\address{Universit\'e Paris Diderot, F-75013 Paris, France\\
CNRS, UMR 7589, LPTHE, F-75005 Paris, France\\
Sorbonne Universit\'es, UPMC Univ Paris 06, UMR 7589, LPTHE, F-75005 Paris, France
\\
cacciari@lpthe.jussieu.fr}

%

\maketitle


\begin{abstract}

We review the history of jets in high energy physics, and describe in more
detail the
developments of the past ten years, discussing new algorithms for jet finding
and their main characteristics, and summarising the status of perturbative
calculations for jet cross sections in hadroproduction. We also describe the
emergence of jet grooming and tagging techniques and their application to
boosted jets analyses.


\keywords{QCD, jets, LHC.}
\end{abstract}



\section{Introduction}

In the context of high energy collisions of elementary particles, `jets' are bunches
of particles that are detected in a contiguous region of a detector, and that stand out
from the rest of the event.\footnote{A precise definition of what exactly `bunches'
and `standing
out' mean is a necessary condition for
performing quantitative physics studies. Formulating such a definition is a 
delicate and potentially complex task, and one of the reasons why a review like this
one is needed to properly address this topic.} The reason why we
are interested in this kind of object is that the
simplest mechanism for producing such bunches is the production of a large momentum
parton, i.e. a quark or a gluon, in the elementary collision, followed by its
quasi-collinear hadronisation and decay into the observable hadrons and leptons. In
this picture a jet is therefore a proxy for the original parton, and studying it allows
one to probe the original elementary collision.

Experimental observation of jets is proof of the asymptotic freedom property of
Quantum Chromodynamics (QCD), the theory of strong interactions: it is the fact that
such interactions become  weak at large energy scales that allows a quark or a
gluon produced at large energy to have its own individuality for a while,
and therefore seed its own jet, before the strong character of the force takes over and
forces it to hadronise. Jets are therefore a `relic' of the first instants after a
collision, at a time $t \ll 10^{-24}$~s, and their study allows one to image the
collision's products at that early time, as if taking a
picture.\footnote{To the author's best knowledge, the first person to liken the study
of jets to taking a picture of a high energy physics event was Gavin Salam in his
`Jetography' review~\cite{Salam:2009jx}.}
In the following we will give a short historical introduction, and we will then
review the modern approach to jets that is being used in predictions and analyses for
the Large Hadron Collider (LHC), as well as the most recent advances.

\section{A short history of jets}

Jets were first used to study the character of the strong interaction force in the 
mid-seventies, when analysing events produced by the $e^+e^-$ collider PETRA at DESY. At the
time QCD had just been put forward as a candidate theory for strong interactions, and
experiments had the task to establish conclusively the existence of the gluon and the
property of asymptotic freedom. In the context of $e^+e^-$ collisions producing, via
electromagnetic interaction, quark-antiquark pairs, this meant observing events with a
structure that could be assimilated to the production of three jets with large energy,
and not only two back-to-back ones, the idea being that a third jet would be seeded by
a gluon, emitted by either the quark or the antiquark. 
 Various algorithms and measures were proposed, and
they are reviewed for instance in Ref.~\citen{Moretti:1998qx}. Suffice here to say that
the `three-jet' character of a (small) fraction of events could be experimentally
established, providing one of the first confirmations of QCD and of asymptotic freedom.

The smallness of the strong coupling in the asymptotic, large-energy regime opens the
way to performing perturbative calculations in QCD to predict the observables that are
measured by experiments. Jet rates can be one such observable, and can therefore
provide not only qualitative but also quantitative insight into QCD, for instance by
allowing one to measure the strong coupling. However, one quickly realizes that in
order to calculate jet-related quantities in perturbative QCD some care must be taken
in defining the jets. 
Like most quantum field theories involving massless particles, QCD develops infrared and
collinear singularities when either an infinitely soft particle is emitted, or when a
particle is emitted collinear to another one. These singularities can be cancelled by
corresponding ones in virtual contributions: this has long known to be the case for
fully inclusive quantities, as shown by the Kinoshita-Lee-Nauenberg
theorem\cite{Kinoshita:1962ur,Lee:1964is}.
However, if an observable, e.g. a jet, 
is not fully inclusive and is not properly defined, 
the cancellation can fail to take place and  the singularities will show up as divergences in the
perturbative result, making it worthless: we will then say that these observables are
infrared and/or collinear (IRC) unsafe. 
It is easy to come up with a very intuitive,
and yet IRC-unsafe definition of a jet. Imagine selecting the three most energetic
particles in an event, and drawing a cone around each of them. Such three jets can be
measured, but their rate cannot be predicted in perturbative QCD: at some order in the
perturbative series one of these three energetic particles can undergo a collinear
splitting into two less energetic particles, such that none of the two now qualifies as
one of the three most energetic ones in the event. This particular event now displays
three different jets at this perturbative order and, from the point of view of the
perturbative calculation, the cancellation of the singularity against the virtual
contributions will not be ensured.

One of the first attempts to define IRC-safe jet cross sections in perturbative QCD 
was made in 1977, when Sterman and Weinberg\cite{Sterman:1977wj} 
spelled out the necessary conditions and provided the
relevant calculation for the next-to-leading order (NLO) correction to a two-jet
rate in $e^+e^-$ collisions. It is worth writing out their result in detail because of
its pedagogical value: using slightly modified notation, the {\sl ``fraction of all
events which have all but a fraction $\epsilon\, [\ll 1]$ of their energy in some pair of 
opposite[ly directed] cones of half-angle
$\delta\,[\ll 1]$''} is given by
\be
f_2 = 
1-\frac{C_F\alpha_s}{\pi}
\left( 4\ln 2\epsilon\ln\delta + 3 \ln\delta + \frac{\pi^2}{3}
-\frac{7}{4} + {\cal O}(\epsilon, \epsilon\ln\delta, \delta^2\ln\epsilon) \right)
\ee
where $C_F=4/3$ and $\alpha_s$ is the strong coupling. In this expression for the two-jet fraction one clearly recognizes 
the `no-emission' limit, $\alpha_s \to 0$, which
implies that the two-jet rate receives no higher order correction, i.e. there is never
a third jet if the probability of radiating a gluon goes to zero. More generally, the
two-jet rate is dominant as long as $\alpha_s$ is small. Moreover, one sees how the definition needs two resolution parameters,
$\epsilon$ and $\delta$, to make it sufficiently inclusive for the real-virtual
cancellation to take place: some radiation is allowed to escape, either because not
sufficiently collinear or because too soft, but only a limited amount. If $\epsilon$
and $\delta$ are chosen too small one probes either very soft or very collinear
radiation, re-exposing the soft and collinear singularities in the form of very large
logarithms in the final perturbative result.

The Sterman-Weinberg jet definition above, while conceptually and historically
important in its opening the way to the definition of IRC-safe observables in QCD, is not the most useful or
practical jet observable that one can think of. For instance, it works with a
predefined number of jets, while often, and even more frequently in high energy
hadronic collisions, the structure of the event can vary greatly, and the task of
finding out how many jets have been produced is better left to an algorithm. Since the
beginning of the Eighties many different such algorithms were proposed. By
`jet algorithm' we mean a set of rules, possibly accompanied by parameters, for
grouping the particles produced in the final state of a high energy collision into
jets. Ideally, no prior assumptions on the structure of the event are made: the
algorithm is simply run over all the particles in it, and it is expected to return the
jets (if any). For this reason jet algorithms are also often called `jet finders'.

Two main classes of jet algorithms soon emerged, the cone-type and the clustering-type
ones. Cone-type algorithms are an evolution of the original Sterman-Weinberg definition:
jets are still expected to be cone-like agglomerations of particles, but the task of
finding how many of them are present in an event and where they are is left to the
algorithm. This was usually accomplished through an iterative procedure, whereby the
flow of energy (i.e. the sum of the four-momenta of the particles) within some cones
was iterated over all placements of the cones, until a
stable situation was reached -- `stable' usually meaning that the momentum sum of the
particles inside a cone coincided with the axis of the cone itself. Cone-type
algorithms are therefore a kind of top-down algorithm: they begin with pre-formed
structures (albeit not necessarily a fixed number of them), and see where these can fit in the
event. The other main class of jet algorithm is instead of bottom-up kind: these
algorithms usually start with the elementary constituents, i.e. the list of the
four-momenta of the particles in the event, and assemble them into larger structures
via an agglomerative clustering procedure by iteratively combining two\footnote{Or more,
see e.g. Ref.~\citen{Lonnblad:1992qd} for an example of an algorithm performing $3\to 2$
 recombinations rather than $2\to 1$ ones.}
four-momenta, until only a certain number is left under a set of stopping rules and
parameters. The clusters that have been formed in this way are then called the jets.

Algorithms from both classes have been used extensively in high energy physics in the
past thirty years, cone-type ones mainly in hadronic collisions and, in the pre-LHC era,
clustering-type ones mainly in $e^+e^-$ collisions. The origin of this split is to be found
in the different characteristics of the two kinds of environments. In hadronic
collisions many particles (from several tens up to, at the LHC, several thousands) are
produced. One must therefore find an algorithm that remains reasonably fast at large
multiplicities. Moreover, the noisy environment of a hadronic collision requires
multiple corrections of experimental kind. These reasons pushed experimentalists to privilege,
e.g. at the Fermilab Tevatron $p\bar p$ collider,  cone algorithms, which could be
implemented efficiently and returned jets with a smooth, often circular profile. In
$e^+e^-$ collisions, instead,  multiplicities are much lower, so that implementation
speed is not a concern, and the clean environment allows for precision measurements and
studies. It turned out that some clustering-type algorithms could be formulated in a
way to allow for all-order resummation in perturbative QCD of some classes of terms,
hence leading to more accurate predictions. Because of this, clustering-type algorithms
were preferably used at the CERN LEP $e^+e^-$ collider. 

It would be impossible to review here, even only superficially, all the specific
algorithms that have been formulated and used in the past. I refer the interested
reader e.g. to Ref.~\citen{Ellis:2007ib} for a review that mainly focuses on hadronic
collisions, to Ref.~\citen{Moretti:1998qx} for one mainly concerned with $e^+e^-$
collisions, and to Ref.~\citen{Salam:2009jx} for a more recent and modern overview.

\section{Jets in the LHC era}

In the run up to LHC in the early 2000's, it became clear that none of the
algorithms that had been and were still in use, at the Tevatron or in preparation
studies for the LHC, was entirely satisfactory.  All cone-type algorithms that were
available at the time were IRC-unsafe, even if some had been patched in order to shift
the unsafety to a higher order in perturbative QCD, as was the case for the MidPoint
algorithm (see Ref.~\citen{Blazey:2000qt} and references therein). The (more or less
severe) IRC unsafety of these algorithms was usually the product of compromises made in
order to keep their running time within usable limits. As a consequence, however, the
high-accuracy NLO and next-to-next-to-leading (NNLO) calculations that the theoretical
community was painstakingly producing for use at the LHC risked facing
measurements for observables for which no finite prediction was possible.
Clustering-type algorithms from the LEP era, on the other hand, were IRC-safe but were
implemented in a way that didn't allow for graceful scaling up to the higher
multiplicities of the LHC, and most of them needed very large experimental corrections.
Hence they were not seriously considered by the experimental community as viable tools
for LHC physics.

This situation began to change in 2005, when a fast implementation of a
clustering-type algorithm, the so-called longitudinally invariant $k_t$
algorithm\cite{Catani:1993hr,Ellis:1993tq}, 
was proposed in Ref.~\citen{Cacciari:2005hq} by G. Salam and this author. This
proposal allowed one to lower the algorithmic complexity of this algorithm from ${\cal
O}(N^3)$ (with $N$ being the number of particles to cluster) to ${\cal O}(N^2)$ and,
making further use of computational geometry techniques, even to ${\cal O}(N\ln N)$.
For an event with $N\simeq 1000$ particles, quite typical at the LHC, this meant
reducing the clustering time on a modern ${\cal O}(1~\mathrm{GHz})$ processor from one
second to one millisecond or less per event, opening the way to practical use of this
IRC-safe algorithm at the LHC.

A few years later, in 2007, G. Salam and G. Soyez put forward the SISCone
algorithm\cite{Salam:2007xv}, a
cone-type algorithm that is IRC-safe while still being implemented in a way that is
sufficiently fast for practical use.
Soon thereafter, however, the pendulum of jet algorithms at the LHC would start to
swing decisively in favour of clustering-type ones.

\subsection{Jet algorithms and their implementations}

In 2008 the following generalisation of the longitudinally invariant $k_t$ algorithm
was presented in Ref.~\citen{Cacciari:2008gp}. One defines the particle-particle and the
particle-beam distances respectively as
\ba
&&d_{ij} = \min(p_{ti}^{2p}, p_{tj}^{2p})\,\frac{(y_i-y_j)^2 + (\phi_i - \phi_j)^2}{R^2}\\
&&d_{iB} = p_{ti}^2
\ea
In these equations $p_t$, $y$ and $\phi$ denote the transverse momentum with respect to
the beam axis, the rapidity and the azimuthal angle respectively of the particles $i$
and $j$. The `jet radius' $R$ and the exponent $p$ are parameters of the algorithm. The
algorithm works by calculating the $d_{ij}$'s and the $d_{iB}$'s for all particles in
the event, and finding the smallest one. If this smallest distance is an inter-particle
one, the two four-momenta are recombined according to a given recombination scheme. If
it is a beam-distance the four-momentum concerned is simply called a jet and excluded
from further processing. The algorithm proceeds in an iterative way, recalculating 
distances after each step, until no particles are left in the event. One then retains all
jets above a given $p_t$ cut.

When the parameter $p$ is set to one, one recovers the longitudinally invariant $k_t$
algorithm of Refs.~\citen{Catani:1993hr,Ellis:1993tq}. The choice $p=0$ gives  an
algorithm, dubbed Cambridge/Aachen\cite{Dokshitzer:1997in,Wobisch:1998wt}, where clustering is based exclusively on the
angular distance between the particles. The distance measures of these two algorithms
are modelled on the physical behaviour of the QCD emission probability. For the $k_t$
algorithm this distance, the
relative transverse momentum of the two particles (hence the $k_t$ name for the
algorithm), is proportional to the inverse of this emission probability. For the Cambridge/Aachen algorithm, a purely angular distance is meant to
mimick the (inverse of) the QCD emission probability in the presence of angular
ordering\cite{Mueller:1981ex,Ermolaev:1981cm}. In both these cases, an algorithm formulated in this way is
expected to roughly `walk back' through the parton-branching process induced by QCD emissions,
and therefore reconstruct approximately the original hard quark or gluon that
fragmented into the final state particles.

Setting $p=-1$ means largely abandoning pretenses of a physically meaningful connection
between the clustering algorithm and the QCD branching process\footnote{This is
the case for momenta, but not for angles: small-angle splittings are still privileged, like in
the  $k_t$ and the Cambridge/Aachen algorithms.}. Perhaps somewhat surprisingly, this turns
out to be a jet algorithm that enjoys many favourable characteristics. It has been
christened `anti-$k_t$' in Ref.~\citen{Cacciari:2008gp}, and it has been adopted as the
default jet algorithm by all the experimental collaborations at the LHC. Among the
useful properties of anti-$k_t$, we mention in particular IRC-safety, and the fact that
it produces jets with very regular borders (often circular, if sufficiently isolated)
and that do not usually extend beyond a distance $\simeq R$ from the hard particle(s)
that seeds them. These latter two properties reduce sensitivity to background noise
(underlying event and pile-up) and facilitate experimental corrections for these and
detector-related effects.

SISCone\cite{Salam:2007xv} is instead an algorithm of cone type, meaning that it is
based on the concept of stable cones, as defined above. Finding all stable cones  can
be a very onerous operation: if performed combinatorially over $N$ particles this
operation has complexity ${\cal O}(N2^N)$, making it totally impractical beyond a
(small) handful of particles. Old cone-type algorithms typically skirted this
problem by resorting to approximate geometrical methods, based for instance on using a
finite number of `seeds' (e.g. some of the particles in the event) as the starting
point of an iterative procedure that searched for stable cones. The problem of these
approximate approaches is that they cannot guarantee that all stable cones will be
found: missing some stable cones means that the algorithm becomes sensitive to
the addition (or removal) of soft or collinear particles, and therefore IRC-unsafe.
The breakthrough of SISCone consists in a geometrical procedure that guarantees that
all stable cones are found, while remaining practical in terms of time taken: $N$
particles can be clustered by SISCone in ${\cal O}(N^2\ln N)$ time. While slower than
the clustering-type algorithms mentioned above, this is fast enough for concrete use at
the LHC.

The $k_t$, Cambridge/Aachen, anti-$k_t$ and SISCone jet algorithms are all implemented
in the FastJet package\cite{Cacciari:2011ma,Cacciari:2005hq}, today's de-facto standard
for jet clustering for LHC phenomenology and experimental analyses. It can be
downloaded from \url{http://fastjet.fr}.

\subsection{Jet areas and background subtraction}

The availability in FastJet of fast implementations of IRC-safe algorithms meant that
one could explore properties of jets that were beforehand either impossible or too
laborious to study. One first such property is the {\sl area} of a jet. Naively, the
area of a jet is the surface, in the rapidity-azimuth plane, covered by the jet itself.
However, since the jet is simply a collection of particles\footnote{This is certainly
the case for clustering-type jets. In the case of cone-type jets an a priori
geometrical interpretation may already be available through the definition of the cones
themselves.}, one cannot immediately determine unambiguously such an area. Jet areas
have phenomenological significance because they can be interpreted as a jet algorithm's
{\sl susceptibility} to contamination from a background of diffuse, roughly uniform
soft radiation in an event. Physical realisations of such a background are pile-up noise
from multiple simultaneous collisions and the underlying event\footnote{To a lesser extent, since it is usually less
uniform and can retain some dependence from the hard collision.}.

\begin{table}
  \centering   
  \tbl{ A summary of main active area results for the four main IRC-safe algorithms 
  $k_t$, Cambridge/Aachen, anti-$k_t$ and SISCone, taken from Refs.~\citen{Cacciari:2008gn}
  and \citen{Cacciari:2008gp}. The values for the active ($A$) areas for
    1-particle jets (1PJ) and for  the magnitude of the active area
    fluctuations ($\Sigma$) are shown, followed by the coefficients of
    the respective anomalous dimensions ($D$ and $S$) in the
    presence of perturbative QCD radiation. All values are normalised to $\pi R^2$, where
    $R$ is the jet radius parameter.}
   {\begin{tabular}{@{}rcccc@{}} \toprule
                  & $A$(1PJ)$/\pi R^2$ & $\Sigma$(1PJ)$/\pi R^2$  &  $D/\pi R^2$ & $S/\pi R^2$       \\\hline 
   $k_t$          & $0.81$   & $0.28$	       & $0.52$   & $0.41$  \\ 
   Cam/Aachen     & $0.81$   & $0.26$	       & $0.08$   & $0.19$  \\ 
   anti-$k_t$     & $1$      & $0$	       & $0$	  &  $0$      \\ 
   SISCone        & $1/4$    & $0$	       & $0.12$   &  $0.07$   \\ 
  \end{tabular}
  \label{tab:summary}}
\end{table}

Jet areas were first studied extensively in a modern context in
Ref.~\citen{Cacciari:2008gn}. In this paper various definitions were given, and the
characteristics of the resulting jet areas were studied analytically and/or numerically.
The main result is the observation that jet areas are not only a characteristic of a
specific jet algorithm, but also depend on the constituents of each jet: jets composed
exclusively of many soft particles will often have typical areas that will differ from
those of jets anchored by a few hard particles. Moreover, jet areas will `evolve' as a
function of the energy scale of the jet, a result of the emission of new particles via QCD
radiation. 
These results can be summarized by the values given in Table~\ref{tab:summary} for the four main
IRC-safe algorithms, $k_t$, Cambridge/Aachen, anti-$k_t$ and SISCone. They express the
average active areas (as defined in   Ref.~\citen{Cacciari:2008gp}) for idealised jets
constituted of a single hard particle  and  filled by a roughly uniform distribution of
much softer radiation. These values can  be interpreted in a physical context as follows:
$k_t$ and Cambridge/Aachen jet areas have similar sizes, on average slightly smaller than
the `circular cone' area $\pi R^2$, but fluctuate a lot from one jet to another. 
Furthermore, jets from these  algorithms have both areas and their fluctuations that
increase at large energy scales,  as shown by the positive coefficients $D$ and $S$ that
control their evolution through the equations\footnote{In these equations $\as$ denotes the strong coupling, $b_0$ the first coefficient of the QCD beta
function, $C_1$ a color factor, $Q_0$ an initial low scale where ideally no extra radiation is present in the jet,
$p_{t}$ the transverse momentum of the original emitting particle, and therefore of the
jet.}\cite{Cacciari:2008gp}
\be
  \langle \Delta A\rangle \simeq
  D \frac{C_1}{\pi b_0}
  \ln \frac{\as({Q_0})}{\as(R p_{t})}\,
\quad\mathrm{and}\quad
  \langle \Delta \Sigma^2\rangle \simeq
  S^2 \frac{C_1}{\pi b_0}
  \ln \frac{\as({Q_0})}{\as(R p_{t})}
  \, .
\ee
The fluctuating and increasing jet areas implied by these equations are the
characteristics that make jets from the $k_t$ or Cambridge/Aachen algorithms difficult to
correct for at the experimental level. On the other hand, one can see from Table
~\ref{tab:summary} that anti-$k_t$ jets containing a single hard particle have instead a
fixed area equal to $\pi R^2$, with zero fluctuations and zero evolution. This result
holds well at the level of physically more realistic jets, explaining why anti-$k_t$ is an 
ideal algorithm to work with at the experimental level and therefore why it has been
chosen as the default algorithm by the LHC collaborations.\footnote{SISCone areas also
have no fluctuations, however they do increase with increasing energy scales. More
importantly, this result does not hold as well as for anti-$k_t$ at the realistic jets
level, where the `split-merge' step of SISCone, needed to deal with overlapping stable
cones, produces jets of different areas and therefore induces fluctuations that are absent
in the ideal 1-particle jet.}

Jet areas can be evaluated for realistic jets using the FastJet package. The possibility
to know, for each jet, its susceptibility to contamination from a roughly uniform
background opens naturally the way to exploit this property to correct for this. In
practice, this can be used to subtract contamination from underlying event and pile-up
radiation effects by subtracting from each jet's transverse momentum a quantity $\rho A$,
where $A$ is a jet's active area and $\rho$ is a transverse momentum density of soft
radiation, per unit rapidity and azimuth, as measured in each event:
\be
p_t^{corrected} = p_t^{raw} - \rho A \, .
\ee
This
procedure\cite{Cacciari:2007fd} successfully corrects a jet's transverse momentum for contamination,
jet-to-jet area fluctuations and event-to-event background level fluctuations, and only
leaves residual uncertainties related to point-to-point fluctuations of the background in
a single event (as one cannot measure it properly inside a hard jet) and back-reaction
effects and fluctuations.\footnote{Back-reaction\cite{Cacciari:2008gp} is related to the
fact  that the constituents of a jet can be modified by whether the clustering is
performed in the presence of neighbouring
particles or not. A further merit of anti-$k_t$ jets is that they are minimally sensitive
to this effect.\cite{Cacciari:2008gp}}

\section{Perturbative predictions for jet cross sections}

Higher order perturbative corrections to jet production are obviously
important for precision phenomenology and measurements. Not only do they give more
accurate predictions for the production cross sections, but they also give a first
peek into the substructure of the jets, that can now be constituted of more than a
single quark or gluon.\footnote{Realistic jets are of course composed of many
constituents as a result of the hadronisation of quarks or gluons, but this
substructure is not under the control of perturbative QCD.}

Among the main processes involving jets and of interest at the LHC we find 
inclusive jet production and the production of a number of jets in association with
an electroweak vector boson or a Higgs boson. We list below the main higher order
perturbative calculations that have been performed in QCD for these processes. For
brevity, we often mention only the reference that appeared first in the
literature\footnote{This is especially true for the NLO calculations, many of which
are quite old and have therefore been repeated a number of times, using different
techniques.}, but it is worth bearing in mind that for calculations of this level of
complexity a second independent result is usually equally valuable, even if it comes
much later.

For inclusive jet hadroproduction, NLO corrections to
one-\cite{Aversa:1989xw,Ellis:1990ek}, two-\cite{Ellis:1992en}, 
three-jet\cite{Nagy:2003tz} rates have been known since some time. Four-\cite{Bern:2011ep}
and five-jet\cite{Badger:2013yda} rates have been calculated to NLO more recently.  No full NNLO calculation for inclusive jet hadroproduction
has yet been completed, but first partial results for the one- and two-jet rates
have appeared in Ref.~\citen{Ridder:2013mf}.

NLO corrections for hadroproduction of an electroweak boson $V$ plus $n$ jets have
 been calculated a long time ago for $n=1$\cite{Giele:1993dj} and
$n=2$\cite{Campbell:2002tg}. 
NLO corrections for $V+n$~jets with $n\ge 3$ have
become available only more recently, thanks to the advent of new techniques that
have greatly simplified the calculation of one-loop amplitudes. They are now known
for $n=3$\cite{Berger:2009zg,KeithEllis:2009bu}, $n=4$\cite{Berger:2010zx} and
$n=5$\cite{Bern:2013gka}.
NNLO corrections for $V+1$~jet have been calculated very recently
in Refs.~\citen{Boughezal:2015dva,Ridder:2015dxa}. 

Finally, Higgs plus one jet is known to NLO\cite{deFlorian:1999zd},  and to
NNLO\cite{Boughezal:2013uia,Chen:2014gva},  while Higgs plus two jets was calculated to NLO
in Ref.~\citen{Campbell:2006xx}  for the gluon-gluon fusion process and in
Ref.~\citen{Figy:2003nv} for the vector boson fusion (VBF) process. $H+3j$ is also known to
NLO for the gluon-gluon fusion process\cite{Cullen:2013saa} and for the VBF
process\cite{Figy:2007kv}.  NNLO corrections to $H+2j$ production for the VBF process
have been presented recently in Ref.~\citen{Cacciari:2015jma}.

\section{Inside the jets: jet substructure}

Clustering-type algorithms of agglomerative kind, like $k_t$, Cambridge/Aachen and
anti-$k_t$, allow one to inspect the {\sl clustering history} of each jet, i.e. the order
in which its constituents were clustered with each other, and at what values of the
distance measure $d_{ij}$ this happened. When this distance measure has physical meaning,
as is the case for $k_t$ or Cambridge/Aachen, this information can be exploited to acquire
knowledge about the splitting processes that originated the constituents of the jets, and
therefore the nature of the initiating particle.

\subsection{Early attempts}

The first attempt to exploit the substructure of clustering-type jets was made by 
M. Seymour in 1993. In Ref.~\citen{Seymour:1993mx} he studied a heavy Higgs decaying into
a pair of $W$ bosons, with one $W$ decaying leptonically and the other hadronically. He
then looked at the two subjets obtained by undoing the last
clustering step of a $k_t$ algorithm with $R=1$: these two subjets should correspond to
the two quarks from the hadronically decaying $W$, since the $d_{ij}$ distance between
them is generally the largest one (because proportional to the $W$ mass\footnote{See e.g.
Section 5.3.1 of Ref.~\citen{Salam:2009jx} for a detailed analytical discussion.}) and therefore 
they 
cluster last in the $k_t$ clustering sequence. Performing this unclustering, and looking in
appropriate invariant mass and angular distance  windows of the two subjets, effectively
tags the $W$ boson. Ref.~\citen{Seymour:1993mx} observed that a better mass resolution
could be obtained in this way rather than directly clustering with cone algorithms with
much smaller radii ($R \sim 0.25$): this latter approach could still allow one to directly
identify the subjets, but the large radiation loss from small jets leads to a poorer mass
resolution.

Further early work that exploited a jet's substructure was performed in
Refs.~\citen{Butterworth:2002tt,Butterworth:2007ke}. In these papers the observation that the distribution of the
$k_t$ distance $d_{ij}$ between the two candidate $W$ subjets is close to the
$W$ mass for real $W$ boson decays, but lower for generic, high-mass QCD jets,
was used to significantly reduce the background. This approach is often
referred to as `Y-splitter'.

\subsection{Renaissance: the BDRS tagger for boosted Higgs}

In a landmark paper\cite{Butterworth:2008iy} published in 2008, Butterworth, Davison, 
Rubin and Salam (BDRS) presented a jet substructure-based analysis for the search at
the LHC of a boosted (i.e. with large transverse momentum $p_t \gg m_H$) Higgs boson
decaying into a bottom-antibottom pair. Such a search faces a huge background from
standard QCD jets, and had previously been deemed unfeasible. BDRS approached the
problem by requiring that the Higgs be highly boosted: this reduces the signal, but
reduces the background even more. Moreover, a boosted Higgs tends to decay in a
collimated way in the laboratory frame, with all decay products to be found within a
single jet with radius $\sim 2 m_H/p_t \simeq 1$ for $p_t > 200$~GeV. This allows one
to devise the following strategy: look for hard `fat' (i.e. large radius)  jets as
possible candidates, and analyse their substructure in order to tag those that exhibit a
two-prong structure, telltale sign of the decay of the Higgs into the $b\bar b$ pair.

\begin{figure}[t]
\centerline{\includegraphics[width=\textwidth]{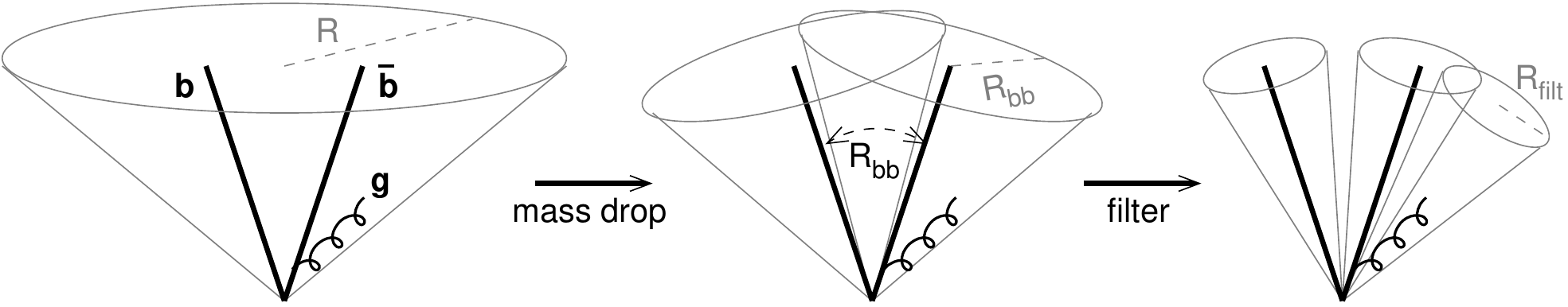}}
\caption{The three stages of the BDRS tagger. Figure reproduced from
Ref.~\citen{Butterworth:2008iy}.
\label{fig:BDRS}}
\end{figure}

In practice, this was achieved in BDRS the following way. An event is first clustered
with a jet algorithm with a large radius $R$. This jet algorithm can be quite generic,
since the constituents of the hard jet of interest can then be reclustered with an
algorithm whose clustering history is physically meaningful and can be exploited to
analyze the jet substructure.  The choice made in BDRS was to use the Cambridge/Aachen
(C/A) algorithm\cite{Dokshitzer:1997in,Wobisch:1998wt} for the reclustering, and therefore exploit the
angular distances of the constituents. The C/A algorithm does not have the property
that the decay products of a massive particle tend to cluster last, as is the case for
the $k_t$ algorithm. BDRS had therefore to devise a way to identify this `relevant
splitting' in the whole clustering sequence. This was done by performing successive
declusterings and, for each step, studying the asymmetry  of the two momenta and  the
`mass drop' of the jet: a declustering step involving not too-asymmetric momenta
and producing a big mass drop is interpreted as having split the jet formed by the two
decay products of the Higgs. One has therefore effectively {\sl tagged} a jet
containing the Higgs.\footnote{In the BDRS paper,  $b$-tags of the two jets into which
the candidate Higgs is expected to have decayed are also required.} On the other hand,
if the tagging conditions are  never met, the jet is interpreted as QCD background.
This tagger (which is also known as the `mass drop tagger') 
was also complemented by an additional refinement, meant to increase
the resolution of the Higgs mass. Dubbed {\sl filtering}, this procedure breaks the
Higgs-tagged jet into smaller subjets of radius $R_{filt}$, and eventually retains
only the three hardest ones. This allows one to retain the decay products of the Higgs
and at most a single gluon emitted by then, and discard much of the underlying event
contamination. A procedure of this kind, meant to `clean' a jet of radiation not pertaining
to the hard process of interest, is often referred to as {\sl grooming}.
The BDRS procedure described above is illustrated schematically in figure
\ref{fig:BDRS}, taken from Ref.~\citen{Butterworth:2008iy}. 

Before closing this section
it is worth adding an observation about the choice made by BDRS of using the C/A
algorithm. This choice forced them to introduce a procedure to identify the `relevant
splitting', while the $k_t$-algorithm would naturally have given this as the last step
in the clustering. However, the advantage of using C/A resides in the fact of working
with angular distances: once a Higgs has been successfully tagged (step 2 in
figure~\ref{fig:BDRS}), the use of C/A means that the radius $R_{bb}$ of the two cones
shown in the figure will be of the same size as the angular distance between them. In
turn, this means that the angular ordering property of QCD ensures that most 
of the radiation emitted by the Higgs decay products will be contained in these two
cones (since it must be emitted at angles smaller than the angle $R_{bb}$ of the
emitting dipole), and one is therefore assured of collecting it all in the end.

\subsection{Taggers and groomers and their experimental validation}

The BDRS paper has spawned a mini-industry of jet substructure-related tools and analyses, with
dozens of tools having been proposed since 2008, and many of them having been validated at
the experimental level. A full review of all these developments is impossible to achieve
here, but we will try to convey the main ideas and describe the most commonly used tools.
To this end, it is worth bearing in mind the main goal of a tagging/grooming procedure: identify jets produced by specific
final states (often boosted heavy particles decaying into hadrons), while discarding
radiation that cannot be traced back to the original particle of interest, so as to improve
resolution in the reconstruction of its mass. 
Two main classes of algorithms have emerged
that can achieve this goal, those based on the {\sl substructure of a jet} (usually probed
through successive declustering of a preexisting jet, or through specifically crafted clustering
algorithms), and those based on the {\sl pattern of radiation} within a jet (often also referred
to as {\sl jet shapes}).

BDRS belongs to the first class, and it achieves its goal by implementing two separate
procedures, mass drop (and asymmetry) criteria during declustering (for tagging), and
filtering (for grooming). Generalisations (and variants) of BDRS-like procedures can also
lead to taggers for three rather than two prongs, to be used for top tagging. Early
examples have been given in Refs.~\citen{Thaler:2008ju,brooijmans,Kaplan:2008ie}, as well
as in Ref.~\citen{Plehn:2009rk}. Two other early algorithms that belong to the jet substructure
class and that have been widely used are {\sl pruning}\cite{Ellis:2009su,Ellis:2009me} and {\sl
trimming}\cite{Krohn:2009th}. 

Pruning works from the bottom up, modifying the way particles
are clustered: at each $1,2\to p$ recombination step, protojets are clustered only if
$\min(p_{t1},p_{t2})/p_{tp} > z_{cut}$ or $\Delta R_{12} < D_{cut}$, and typical values
for the parameters are $z_{cut} \simeq 0.1$ and $D_{cut} \simeq 1$. Pruning tries
therefore to veto recombinations where either one of the particles is much softer than the
other, or there is a large angular distance between the particles, properties that can characterise either soft
radiation, or emissions from unrelated processes (e.g. underlying event or pileup).
Complemented with a procedure to tag a pruned jet that can originate from a heavy
particle, e.g. if after pruning its mass is within a given window, pruning can act at the
same time as a tagger and a groomer.

Trimming works instead from the top down, taking a jet and splitting it into a number of
smaller subjets of smaller radius. Of these subjets only those with a transverse momentum
larger than a certain fraction $f_{cut}$ of the full jet transverse momentum are retained,
and are then recombined into a `trimmed' jet. One can see that trimming works in a way which is 
very
similar to filtering\cite{Butterworth:2008iy}, differing in using a cut on the
subjets'  momentum rather than on selecting a number of the hardest ones, and like pruning
it can act at the same time as a tagger and a groomer if supplemented with a tagging
criterion. 

We mentioned above that a second class of taggers is based on the study of the patter of
radiation within a jet. The most widely used member of this class is probably {\sl
$N$-subjettiness}\cite{Thaler:2010tr}. This jet shape is constructed as follows. 
Given a jet $J$, $N$ candidate subjets are identified\footnote{The $N$ subjets could be
determined by minimizing $\tau_N$ over all possible subjet directions or, in a
computationally less intensive manner, by running an exclusive $k_t$ algorithm and forcing
it to return exactly $N$ jets, as done in Ref.~\citen{Thaler:2010tr}.}, and the jet shape variable 
$\tau_N$
is then calculated as
\be
\tau_N = \frac{1}{R\sum_{k\in J} p_{tk}} \sum_{k\in J} p_{tk} \min_{i\in subjets}\left\{\Delta R_{ik}
\right\}\, ,
\ee
where the $p_{tk}$ are the transverse momenta of all constituents of the jet $J$, and the
$\Delta R_{ik}$ represent the distance in the rapidity-azimuth plane between the
constituent $k$ and the subjet $i$. $R$ is the original jet radius. One can
 see from the definition above that the values of $\tau_N$ will be
distributed between zero and one. It is moreover readily apparent that $\tau_N$
will be minimized (i.e. $\tau_N \approx 0$) when all radiation is aligned with
the $N$ candidate subjets directions, whereas large values of $\tau_N$ will
signal that there are at least $N+1$ subjets in the jet. These considerations
suggest that the value of the {\sl ratio} $\tau_{N}/\tau_{N-1}$ is a good variable
to discriminate jets with at least $N$ subjets from background jets with $N-1$ or
fewer subjets. In practice, $\tau_2/\tau_1$ and  $\tau_3/\tau_2$ ratios are
usually employed to tag boosted $W$ bosons (two-prong decays) and boosted top
quarks (three-prong decays) respectively.

Grooming techniques like filtering, pruning and trimming, as well as taggers
(both of the jet-substructure and of the  radiation-pattern kind) for
two- and three-prongs, have been extensively validated  on data by some
LHC experiments, notably ATLAS\cite{Aad:2013gja} and
CMS\cite{Chatrchyan:2013rla,Khachatryan:2014vla}, and are now being used in
measurements and searches.

\subsection{Maturity: theoretical understanding and recent developments}

The taggers and groomers described above, mass drop in BDRS,
trimming and pruning, as well as many others, were largely
developed with the help of Monte Carlo event generators like PYTHIA and HERWIG.
Initial theoretical insight was tested using simulated but realistic events,
the performance of the tool was assessed and improved through an iterative process,
and the results could be compared to those of other approaches. This process has led
to considerable understanding of the behaviour of the many taggers and groomers
introduced in the past few years. However, it also has shortcomings. The
performances that are observed in this way could depend critically on any one of
the many details of the simulation of the Monte Carlo, possibly preventing a full
understanding of what really matters and hindering further refinements. Perhaps
more importantly, Monte Carlo simulations can be very time consuming, and often one
cannot fully explore the parameters' phase space: an unexpected behaviour of
a tool for a given choice of parameters may therefore be missed. Because of this,
some recent papers, e.g. Refs.~\citen{Feige:2012vc,Dasgupta:2013ihk,Larkoski:2014wba}, 
have started to analyse the behaviour of jet substructure tools
from first principles in QCD, i.e. using analytical (resummed) perturbative
calculations.  

An explicit example of the importance of a proper theoretical understanding of
taggers and groomers is given by the analysis of the mass drop tagger,
trimming and pruning performed in Ref.~\citen{Dasgupta:2013ihk}. This paper showed
that the behaviour of the three tools as observed in Monte Carlo simulations could
be reproduced to a very good extent using analytical calculation, hence ensuring
an optimal understanding.  Non-perturbative effects could be
studied, and in some cases the analytical results even suggested
adjustments to the design of the tools. 
Perhaps more importantly from the phenomenological point of view, the
three tools could also be easily studied in a wider region of their parameters'  
phase space. It then
became readily apparent that, while the three tools perform quite similarly in
terms of suppression of background QCD jets with large transverse momentum ($p_t
\sim 3$~TeV) and with fairly large invariant masses,  $m \gsim 300$~GeV, important
differences can arise for smaller masses, leading to a potential
background-shaping effect for trimming and pruning. While previous Monte Carlo
simulations had never been extended into this region, the analytical study was
able to easily uncover this effect.

As a second example of the role of analytical calculations in the study and
development of jet substructure tools, we wish to mention `soft drop
declustering'\cite{Larkoski:2014wba}. This procedure has been designed and
engineered from the very beginning with the help of resummed perturbative
calculations. It is a tagger/groomer that aims to remove soft and large-angle
radiation, and generalises and simplifies\footnote{In a perhaps somewhat counterintuitive
way, the mass drop tagger is simplified in `soft drop' by removing the ... mass drop
condition.}  the mass drop tagger: given a jet of
radius $R$ with two constituents, soft drop removes the softer
one unless
\be
\frac{\min(p_{t1},p_{t2})}{p_{t1}+p_{t2}} > z_{\mathrm{cut}}
\left(\frac{\Delta R_{12}}{R_0}\right)^\beta \, .
\ee
This procedure can be applied in a recursive way while successively 
declustering jets with more than two constituents. $z_{\mathrm{cut}}$ and $\beta$
are parameters of the algorithm, and Ref.~\citen{Larkoski:2014wba} has studied
analytically their dependence in a number of different observables. An interesting
characteristic of `soft drop' is that, according to the value of $\beta$, it can
behave either as a groomer (for $\beta > 0$, which only removes soft radiation) or
a tagger (for $\beta \le 0$, which also removes soft-collinear radiation, leading to
the tagging of two-prong decays). For $\beta=0$ soft drop behaves like the (modified) mass
drop tagger introduced in Ref.~\citen{Dasgupta:2013ihk}.

\section{Conclusions}

The field of jet physics has seen much progress in the past ten years.
Standardised algorithms, infrared and collinear safe, are now used by all
experimental collaborations, facilitating comparisons of measurements and allowing
for meaningful comparisons with theoretical calculations. Beyond the jets as
observables, their substructure and the pattern of radiation within them are now
used to make the analyses of final states in high energy collisions ever more
powerful: tagging of massive objects producing boosted jets, and subtraction of
soft-radiation contamination from underlying event and pileup, are prime examples.

Among the recent developments, the shift away from `design by trial and error'
using Monte Carlo event generators, and towards `design by theoretical
understanding' that exploits first principle calculations in QCD, promises to
advance the field even further in the coming years.

The exploitation of the LHC in the coming decade will certainly see no lack of new
jet-based tools at the disposal of phenomenologists and experimentalists alike, in
the quest for ever more precise measurements and, hopefully, discoveries.

\vspace{.5cm}
\noindent

{\bf Acknowledgments.} This review has been made possible by the ten years that I have
spent working on jets with Gavin Salam and Gregory Soyez, in a collaboration that has been
very  enriching for me both on the professional and the personal level. Without them I
would certainly not have been part of the exciting developments that have taken place in
the past few years, and that will certainly accompany us during the exploitation of the LHC
and beyond. I also wish to thank Gavin Salam for a careful reading of the manuscript
and for comments, and Giulia Zanderighi for useful discussions about the higher order
perturbative calculations for jet cross sections.

\end{document}